\documentclass[%
 aip,
 amsmath,amssymb,
 reprint,%
]{revtex4-1}
\usepackage{graphicx}
\usepackage{dcolumn}
\usepackage{bm}
\usepackage{amsmath}
\usepackage{xcolor}

\begin{document}


\title{Plasmonic-cavity Modulator for the Mid-IR with a Semi-transparent and Nonlinear Heavily-doped Semiconductor Mirror}


\author{Tommaso Venanzi}
\altaffiliation{These authors contributed equally to this work.}
\affiliation{Center for Life Nano- and Neuro-Science, Istituto Italiano di Tecnologia, 00161 Rome, Italy}
\author{Raffaella Polito}
\altaffiliation{These authors contributed equally to this work.}
\affiliation{Istituto di Fotonica e Nanotecnologie, Consiglio Nazionale delle Ricerche, 00133 Rome, Italy}
\author{Andrea Rossetti}
\affiliation{Department of Physics and Materials Science, University of Luxembourg, L-1511 Luxembourg, Luxembourg}
\author{Markus Ludwig}
\affiliation{University of Luxembourg, 162a Avenue de la Faïencerie, L-1511 Luxembourg}
\affiliation{Institute for Advanced Studies, University of Luxembourg Campus Belval, L-4365 Esch-sur-Alzette, Luxembourg}
\author{Daniele Brida}
\affiliation{Department of Physics and Materials Science, University of Luxembourg, L-1511 Luxembourg, Luxembourg}
\author{Adel Bousseksou}
\affiliation{Centre de Nanosciences et de Nanotechnologies (C2N), CNRS UMR 9001, Universit´e Paris-Saclay, 91120 Palaiseau, France}
\author{Isabelle Sagnes}
\affiliation{Centre de Nanosciences et de Nanotechnologies (C2N), CNRS UMR 9001, Universit´e Paris-Saclay, 91120 Palaiseau, France}
\author{Gregoire Beaudoin}
\affiliation{Centre de Nanosciences et de Nanotechnologies (C2N), CNRS UMR 9001, Universit´e Paris-Saclay, 91120 Palaiseau, France}
\author{Raffaele Colombelli}
\affiliation{Centre de Nanosciences et de Nanotechnologies (C2N), CNRS UMR 9001, Universit´e Paris-Saclay, 91120 Palaiseau, France}
\author{Antonio Valletta}
\affiliation{Istituto per la Microelettronica e Microsistemi, Consiglio Nazionale delle Ricerche, 00133 Rome, Italy}
\author{Andrea Notargiacomo}
\affiliation{Istituto di Fotonica e Nanotecnologie, Consiglio Nazionale delle Ricerche, 00133 Rome, Italy}
\author{Francesco Mattioli}
\affiliation{Istituto di Fotonica e Nanotecnologie, Consiglio Nazionale delle Ricerche, 00133 Rome, Italy}
\author{Gonzalo Álvarez-Pérez}
\affiliation{Center for Biomolecular Nanotechnologies, Istituto Italiano di Tecnologia, 73010 Arnesano, Italy}
\author{Cristian Ciracì}
\affiliation{Center for Biomolecular Nanotechnologies, Istituto Italiano di Tecnologia, 73010 Arnesano, Italy}
\affiliation{Neurophos Inc., 4120 Freidrich Ln, Austin, Texas 78744, USA}
\author{Valeria Giliberti}
\affiliation{Department of Physics, Sapienza University of Rome, 00185 Rome, Italy}
\affiliation{Center for Life Nano- and Neuro-Science, Istituto Italiano di Tecnologia, 00161 Rome, Italy}
\author{Marialilia Pea}
\affiliation{Istituto di Fotonica e Nanotecnologie, Consiglio Nazionale delle Ricerche, 00133 Rome, Italy}
\author{Michele Ortolani}
\affiliation{Department of Physics, Sapienza University of Rome, 00185 Rome, Italy}
\author{Huatian Hu}
 \email{huatian.hu@iit.it}
\affiliation{Center for Biomolecular Nanotechnologies, Istituto Italiano di Tecnologia, 73010 Arnesano, Italy}
\affiliation{POLIMA—Center for Polariton-driven Light–Matter Interactions, University of Southern Denmark, Odense M, Denmark}


\date{\today}

\begin{abstract}

We present a free-space plasmonic modulator based on a single heavily-doped semiconductor layer. We investigate its ability to modulate both the linear and nonlinear response at mid-infrared frequencies slightly below the plasma frequency of the semiconductor. The device employs a metal–oxide–semiconductor field-effect structure to electrically tune the plasmonic resonances of 1D-patch cavities. Each cavity is defined by a metal grating and a doped semiconductor layer that acts as a tunable mid-infrared mirror, with the gate oxide serving as the dielectric spacer. The gate voltage actively tunes the free-electron density at the semiconductor-oxide interface, where the resonant plasmonic near-field is also confined. We demonstrate electric control of the linear transmittance and reflectance, and of the efficiency of third-harmonic generation. We discuss further performance optimization of the device in terms of modulation speed and depth towards a fast modulator with very simple active material requirements. Our results establish a viable route toward practical plasmonic modulators and mixers operating in the mid-infrared atmospheric window available for free-space communications at wavelengths between 8 and 12 $\mu$m.

\end{abstract}


\maketitle

\section{Introduction}

Active electrical control of resonant optical nanostructures represents a promising approach for optical communication and optical computing functions such as optical modulation, beam steering, multiplexing, mixing and equalization. In optical modulator research, the goals are to enhance the modulation speed by reducing the active device area, and therefore the device capacitance, and to ease the external control parameter, for example from sophisticated (analog) acoustic waves or magnetic fields, currently used in acousto-optic or magneto-optic modulators, to simpler (digital) electric voltage signals that impose free carrier density modulations in integrated waveguides \cite{hiraki2017}. In the last years, plasmonic-resonance modulators have attracted considerable interest as they can feature reduced device areas while maintaining free-space coupling \cite{smolyaninov2019}. In the mid-infrared (IR) range, modulators are useful for applications such as laser amplitude/frequency stabilization, spectroscopy and sensing, and optical communications. However, they exhibit even more severe device area constraints and waveguide-integration difficulties \cite{grillot2025progress}. Plasmonic modulators based on the Stark effect in semiconductor quantum wells have been shown to attain excellent modulation depths \cite{BelkinLinear} and strong-coupling micro-cavity modulators have demonstrated high modulation speed beyond 10 GHz \cite{Colombelli2024, didier2026thin}. We have recently shown that hydrodynamic free electron oscillations in mid-IR plasmonic resonators are inherently nonlinear with giant effective nonlinear coefficients \cite{rossetti2025control}, therefore tuning the electron density therein would lead not only to linear modulators, but also to the active control of the nonlinearity \cite{DelucaPRL, DelucaPRB, hu2024low, alvarez2025ultrahigh, hu2025modulating}. The latter function has been recently obtained in gated quantum-well heterostructures \cite{BelkinTHG,BelkinNonlinear}, but our free-electron modulator would feature much simpler material-growth requirements.

Conventional plasmonic materials (i.e. noble metals) are difficult to modulate by means of a field-effect gate contact, because they display the highest free-electron concentrations in natural materials. To overcome this problem, extremely thin metal films or 2D semimetals have been shown to work effectively \cite{ayata2017high, maniyara2019tunable, yao2018broadband}, but they present challenges in the material growth and they require a proper quantum modeling. Alternatively, if mid-IR wavelengths are targeted, artificial materials with lower electron densities can be employed, such as semiconductor quantum wells \cite{TredicucciACQW, kuo2005strong, pirotta2021fast} and graphene \cite{chen2012optical, fei2012gate, CapassoGraphene}, which show very strong carrier density tunability at low gate voltages. An intermediate option with easier material science challenges and lower requirements for industrial scale production is represented by \textit{not-so-thin} (order $\sim 10^1 - 10^2$ nm) layers of heavily-doped semiconductors with small electron effective mass $m^* \sim 0.05 - 0.15 \mathrm{m_e}$ such as InGaAs, Ge, InP, InGaSb. For a relatively low equilibrium carrier density $ n\sim 10^{18} - 10^{19}\,\mathrm{cm}^{-3}$ these materials display both mid-IR plasma resonances at wavelengths down to 3 $\mu$m \cite{Frigerio, Prucnal, Law12, FischerActivation, rossetti2025control, Panah16} and a significant electron-density tunability in metal-oxide-semiconductor (MOS) field-effect configurations, which are generally available for these technology-relevant semiconductor materials. Heavily doped semiconductors have already shown good optical transmission performance in the near-IR \cite{arcangeli2016gate,tao2019gate,ghindani2021gate}, where plasmonic modulators have already been demonstrated \cite{dionne2009plasmostor}. 

Mid-IR linear and nonlinear modulators are an active topic of research \cite{montesinos2022mid,li2019ge,nedeljkovic2011free, pirotta2021fast} as key figures of merit of modulators have not yet been optimized simultaneously, such as the modulation speed, depth and bandwidth, and the control of optical nonlinearity. To date, free-space mid-IR modulators with the potential to operate up to GHz frequencies in transmission mode, and providing amplitude modulation in both the linear and nonlinear regimes, are still missing. The device presented in this work is analogous to 1D patch cavities that have shown extremely high light-matter interaction levels \cite{todorov2009strong, greffet2023}, including for the generation of nonlinear effects \cite{yu2025full, hu2024low, hu2025modulating}. The main novelty is that the bottom metal ground plane is substituted by the heavily-doped semiconductor layer with plasma frequency in the mid-IR, and a MOS structure is implemented for electrical tunability of the free electron density therein. The use of a heavily doped semiconductor close to its plasma frequency offers the further advantage of a semi-transparent ground plane, which can be exploited to operate the modulator in transmission mode, at odds with existing patch cavity modulators that operate in reflection mode only \cite{pirotta2021fast, BelkinTHG, BelkinLinear, BelkinNonlinear, yu2025full, todorov2009strong}. In our architecture, the 1D patch array plays a dual role: on the one hand, it enables the excitation of localized resonant plasmonic modes close to the mid-IR plasma frequency of the semiconductor layer; on the other hand, it acts as grating-gate electrode, allowing the carrier density to be modulated via the field-effect within the first few nanometers of the doped semiconductor layer. Crucially, owing to the strong confinement of both the plasmonic field-enhancement and the gate-induced band-bending at the oxide-semiconductor interface, the spatial region where the charge density is modulated electrically corresponds to the region where the local dielectric function influences the plasmonic mode frequency, bandwidth, and magnitude \cite{DelucaPRL, hu2025modulating}. This spatial overlap ensures an efficient electrical control of the plasmonic resonances and represents the core concept of our work. 


\section{Modulating free-electron gases in heavily doped semiconductor}

\begin{figure*}
    \centering
    \includegraphics[width=0.8\textwidth]{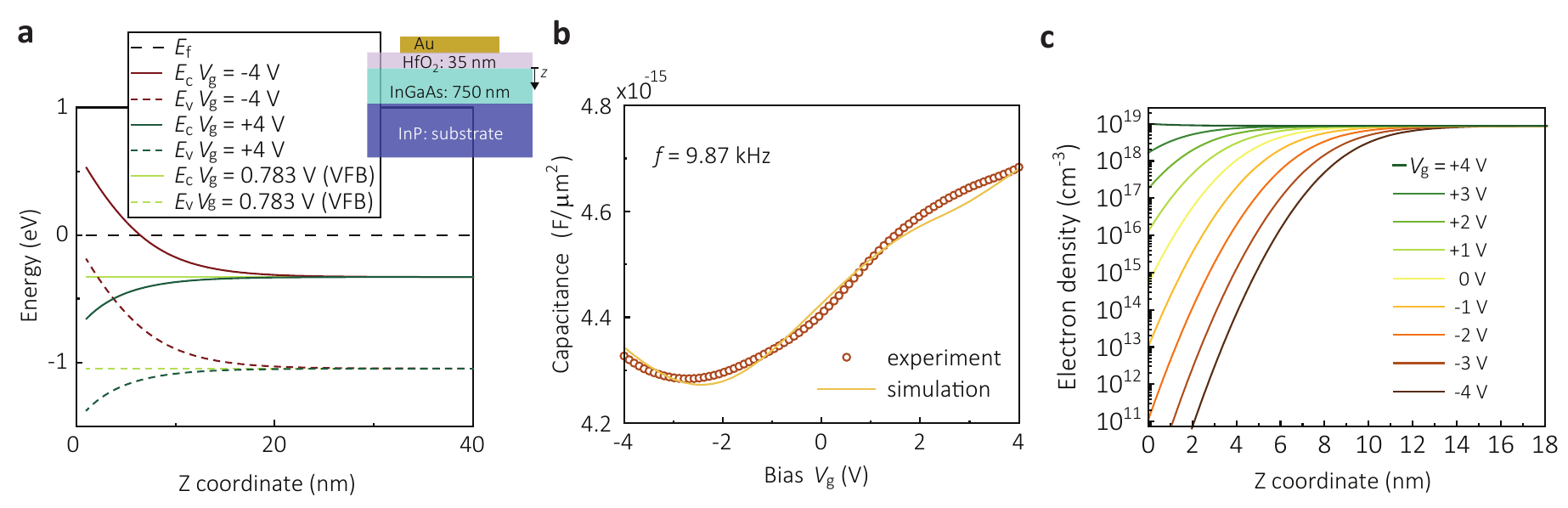}
    \caption{\textbf{Field-effect structure in InGaAs/HfO$_2$.} (a) Conduction (C) and valence (V) band diagram computed near the The HfO$_2$/InGaAs interface of the MOS structure in accumulation ($V_g= +4$ V), inversion ($V_g= -4$ V) and flat band condition (curves labelled with “VFB”, $V_g= 0.783$V). The HfO$_2$/InGaAs interface is located at z coordinate = 0 $\mu$m. The dashed line shows the position of the Fermi level. Inset: schematic of the MOS structure layer stack). (b) Experimental CV characteristic (dotted line) of the MOS structure and simulated CV characteristic (solid line) including a distribution of interface states. (c) Electron density profiles for different values of $V_g$ extracted from simulations. 
    }      
    \label{fig:1}
\end{figure*}

The schematic of the metal-oxide-semiconductor (MOS) structure investigated in this work is shown in the top-right inset in Fig. \ref{fig:1}(a). The gate contact consists of Ti (10 nm)/Au (50 nm) evaporated on a 35 nm-thick HfO$_2$ layer, which serves as the gate insulator. The HfO$_2$ layer is deposited by Atomic Layer Deposition (ALD) on a 750 nm-thick, heavily n-doped In$_{0.52}$Ga$_{0.47}$As layer (doping density $N_D=8.8\times10^{18}$ cm$^{-3}$), grown by Metal Organic Chemical Vapor Deposition (MOCVD) on a slightly doped InP substrate ($N_D=10^{17}$ cm$^{-3}$). Fig. \ref{fig:1}(a) displays the details of the computed band diagram (for both conduction (C) and valence (V) band) near the insulator/semiconductor interface for three different gate conditions: accumulation ($V_g= +4$ V), inversion ($V_g= -4$ V) and a gate voltage level that results in flat band condition ($V_{FB}=0.783$ V).
Because of the high doping level, the InGaAs layer is degenerate with the Fermi level (black dashed line) lying above the conduction band edge at zero bias, and the band bending is confined within the first few nanometers of the semiconductor. The $\pm4$ V bias values were selected as they correspond to the maximum ones employed in this work and define the safe operating window ensuring leakage-free behavior of the gate insulator.

\begin{figure*}
    \centering
    \includegraphics[width=0.7\textwidth]{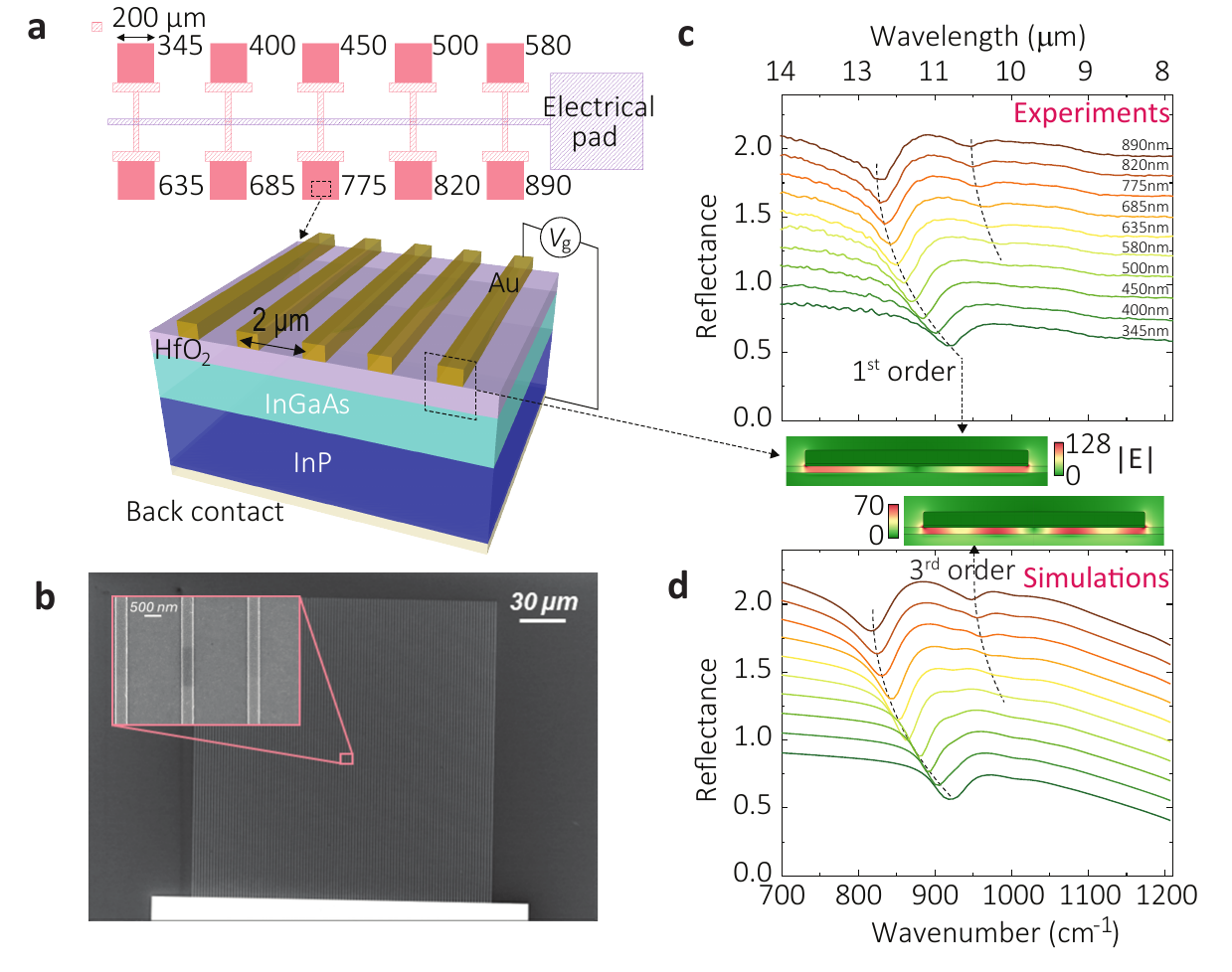}
    \caption{\textbf{Metal gratings on InGaAs/HfO$_2$.} (a) Device layout comprising ten gratings with pitch $p=2000$ nm and varying finger width $w$, all connected to a single electrical pad. A 3D sketch of the layer stack (HfO$_2$/InGaAs/InP) is also shown with the grating-gate structure on top and backside ohmic contact. (b) SEM images with two different magnifications of the grating-gate with $w=$ 345 nm.  (c) Infrared extinction experimental spectra and (d) numerical simulations, also showing the field distribution of the two main plasmonic modes, enhanced below the grating fingers, so as to get optimal spatial overlap with the gate-depleted region of InGaAs at the interface with the HfO$_2$ gate oxide, where the field also extends. Spectra in panels (c) and (d) for $w > 345$ nm have been offset by $+0.1$ each for clarity.
    }  
    \label{fig:2}
\end{figure*}

To monitor the field-effect charge density modulation at the semiconductor/insulator interface, capacitance-voltage (CV) characteristics were acquired on the MOS structure. A Ni (50 nm)/Au (50 nm) back-side ohmic contact was deposited as the reference electrode. Fig. \ref{fig:1}(b) reports the experimental CV characteristics (dots), acquired at an AC frequency of 9.87 kHz, revealing a clear gate-dependent capacitance modulation that reflects carrier density variations in the InGaAs layer at the interface with HfO$_2$. The experimental features are well reproduced by numerical simulations (solid line in Fig. \ref{fig:1}(b)) by including a fitted distribution of interface density of states (DOS) at the semiconductor/insulator interface. Technical details on the numerical simulations can be found in the Supplementary Material (SM). From the simulations, the electron density profiles near the insulator/semiconductor interface were extracted for different gate bias values and are reported in Fig. \ref{fig:1}(c). Notably, due to the introduction of interface states, the n-doped InGaAs is in a depletion condition even for $V_g = 0$ V and reaches a maximum depletion width of the order of 6 nm at $V_g= -4$ V.

\section{Gate-modulation of the plasmonic response}

\begin{figure*}
    \centering
    \includegraphics[width=0.8\textwidth]{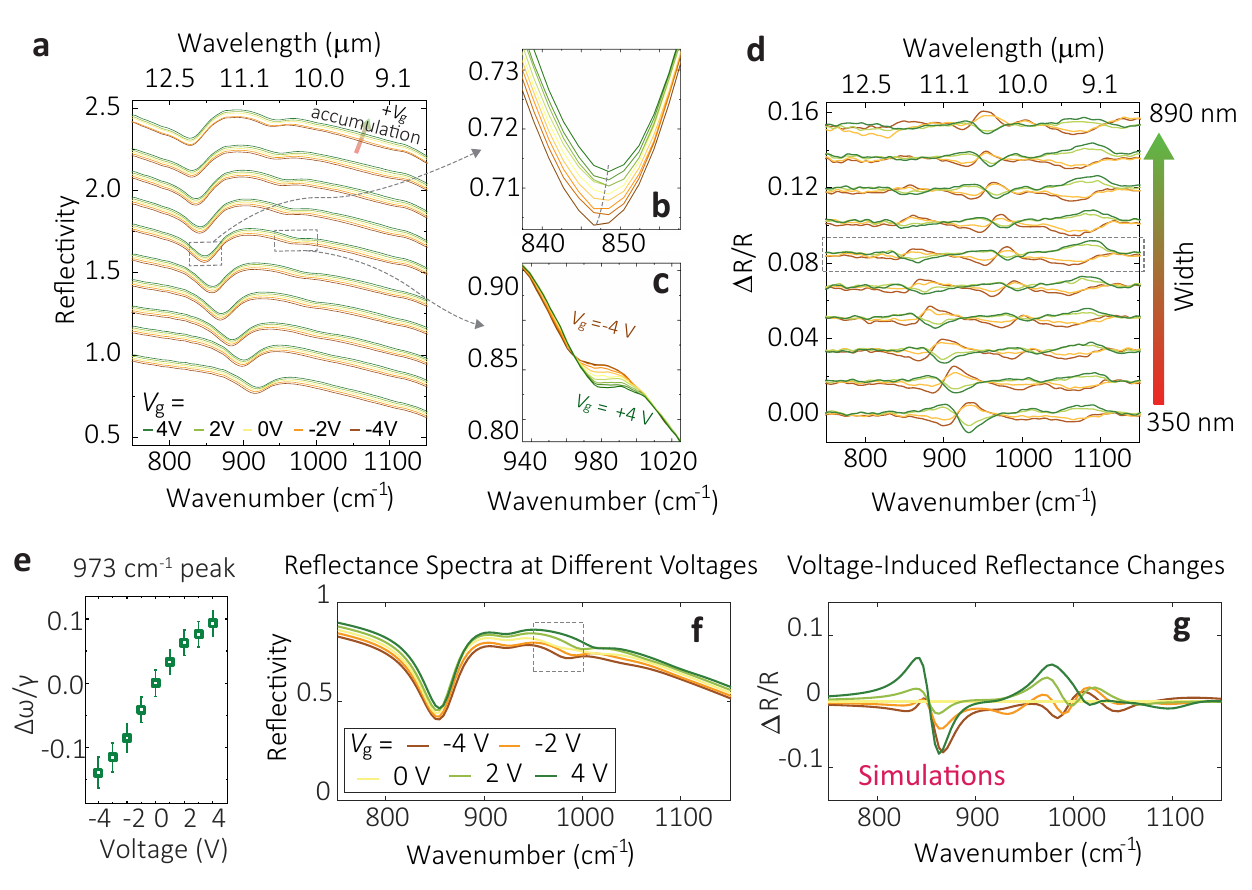}
    \caption{\textbf{Gate-modulation of the linear response.} (a) Reflectance spectra at different applied voltages. The curves are offset for clarity. (b, c) Zoom-in of the dependence of $\omega_{\mathrm{res}}(V_g)$ for $w = 635$ nm. The curves are vertically offset with a different offset value than in panel a. (d) Differential spectra at different applied gate for all the grating finger widths. (e) Relative frequency shift $\Delta\omega$ over the linewidth $\gamma$ as a function of applied voltage. (f, g) Numerical simulations of the differential spectra, corresponding to one case of (d) marked in the dashed box.
    }   
    \label{fig:3}
\end{figure*}

To achieve gate-tunable and optically-accessible plasmonic resonances, we design a metal grating structure that forms 1D-patch cavities filled with the insulating HfO$_2$ layer and using the heavily doped-InGaAs as the ground plane. The schematic layout of the grating-gate pattern is presented in Fig. \ref{fig:2}(a) with a three-dimensional (3D) sketch of a representative cross-section of the stack. The pattern comprises ten gratings, each formed by arrays of rectangular gate electrodes covering a total area of 200 $\mu$m $\times$ 200 $\mu$m. The grating pitch is fixed at $p = 2 \;\mu$m, while the 1D patch width $w$, measured by scanning electron microscopy (SEM), spans the range 345 - 890 nm. All gratings are electrically connected to a single electrical pad. The backside metallic contact was deposited exclusively at the chip edges to permit optical transmission through the grating-gate region. Two SEM images of the 345 nm grating with different magnifications are shown in Fig. \ref{fig:2}(b). Note the small footprint of the metal patch leading to a low internal capacitance, which could be exploited in a RF-compatible device design for GHz-range modulation rates.

The fingers serve both as gate electrodes and as the top mirror of one-dimensional patch cavities. It should be emphasized that the heavily doped InGaAs underneath has a plasma wavelength of $\lambda_p=7.97~\mu$m, thereby acting as a  quasi-metallic mirror for wavelengths above $\lambda_p$. In this configuration, the electromagnetic field is enhanced inside the cavities for specific resonance frequencies. The optical field-enhancement and the DC field-effect regions are guaranteed to effectively overlap with each other, enabling efficient modulation of the plasmonic resonances.

\begin{figure*}
    \centering
    \includegraphics[width=1.0\textwidth]{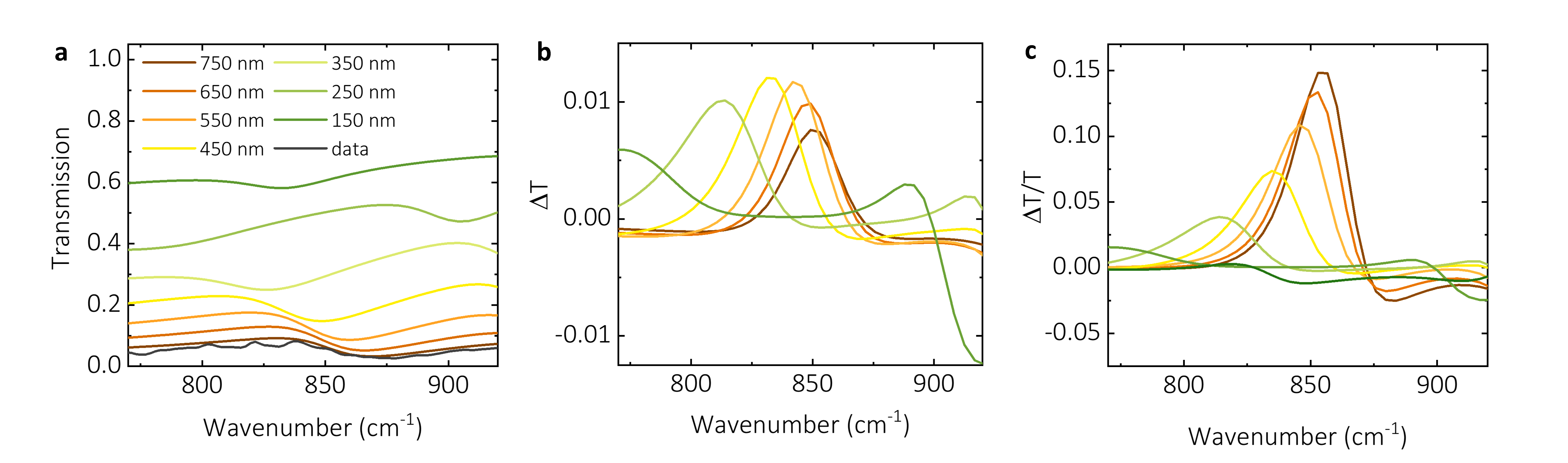}
    \caption{\textbf{InGaAs thickness dependence of the device transmission and modulation.} (a) Simulated transmission spectra for various InGaAs layer thicknesses. The black curve is the experimental transmission of our device, which has a thickness of 750 nm. (b) and (c) Calculated absolute and relative transmission changes for devices with different $d$.
}   
    \label{fig:A6}
\end{figure*}

Figure \ref{fig:2}c shows the infrared reflectance spectra of each grating measured by Fourier-Transform Infrared (FTIR) spectroscopy with an infrared microscope (further details in the SM). We observe two resonances that red-shift with increasing $w$ on top of a broad, high-reflectance baseline increasing from 0.6 at around 1200 cm$^{-1}$ to 0.9 at 700 cm$^{-1}$, as expected for frequencies lower than, but close to, the plasma edge of the n-doped InGaAs ($7.97\;\mu$m$\, \approx\, 1/1255 $ cm$^{-1}$ ). 

To elucidate the origin of the two resonances, we perform electromagnetic finite-element calculations (see SM for details). The two resonances are the first and third-order plasmonic modes of the metal-semiconductor 1D-cavity, revealed by the number of nodes in the mode profiles (Fig. \ref{fig:2}d). For both resonances, the electric field is concentrated in the HfO$_2$ dielectric layer between the metal finger and the doped semiconductor, as expected in a patch cavity. Both modes are localized plasmons with radiation field enhancement around 130-fold and 70-fold, respectively. The regions of the semiconductor not covered by the metal fingers do not significantly contribute to the plasmonic resonances. 

We observe a saturation of the shift in frequency of the plasmonic mode both in the experimental and calculated spectra \ref{fig:2}c and d. The resonances of wider fingers from 770~nm to 890~nm show avoided-crossing at $\approx850$~cm$^{-1}$ that corresponds to the optical phonon band in HfO$_2$ \cite{franta2015universal}. Using the dispersive refractive index reported by Franta \textit{et al.} \cite{franta2015universal}, the observed $\omega_{\mathrm{res}}(w)$, which is defined as the frequency of the local minimum in the reflectance spectrum, is well reproduced in the simulations, whereas an inverse linear scaling $\omega_{\mathrm{res}}(w) \propto 1/w$ using a non-dispersive permittivity would not reproduce the data. 

We next investigate the gate tunability of the linear optical response. Figure \ref{fig:3}a-c shows the reflectance spectra as a function of the applied gate voltage from-4 V to +4 V for all gratings. As observed in Figures \ref{fig:3}b-c, the field-effect charge modulation effectively shifts $\omega_{\mathrm{res}}$. To quantitatively estimate the modulation of the device, we fit the reflection spectra starting from a Drude-Lorentz model with two additional Lorentzian functions to take into account the Drude peak and two plasmonic resonances. The reflection is then calculated using Fresnel equations and assuming a zero degree angle of incidence. As shown in Figures \ref{fig:3}c and e, we find that the third-order peak shifts by about 10 cm$^{-1}$ (30\% of its linewidth).

Figure \ref{fig:3}d shows the differential reflectance with respect to the reflectance at zero applied bias. The resulting spectra exhibit a symmetric dispersive lineshape, indicating a shift of the plasmonic resonances \cite{temperini2024infrared}. The modulation produces a redshift or a blueshift for negative or positive $V_g$, respectively. Moreover, the modulation efficiency depends on the finger width. Gratings with smaller finger widths exhibit a strong modulation of the first-order plasmonic resonance and a weaker modulation of the third-order resonance. In addition to the modulation of the two main plasmonic modes, we observe the modulation of the reflectance baseline peaking around 1080 cm$^{-1}$. 

The modulation of the infrared linear response is well captured by our numerical simulations. In the electromagnetic model, we account for the induced charge density profiles as obtained from the electrical characterization in Figure \ref{fig:1}(c). The results of the simulation are shown in Figures \ref{fig:3}(f-g) for $w=$685 nm. The calculation predicts a redshift of the two peaks that we observe in the experiment. The good  agreement with the $V_g$-dependent experimental data provides useful effective parameters for the future optimization of the device design, which we briefly introduce below. As said, the maximum modulation depth of about 2\% is found at the first-order $\omega_{\mathrm{res}}$. This small value is due to the high-reflectance baseline. To increase the modulation depth, a thinner InGaAs layer thickness $d$ is required to reduce reflection and simultaneously increase transmission. In Figure \ref{fig:A6}, we show the dependence on $d$ of transmittance $T$, modulation $\Delta T$ and relative modulation $\Delta T/T$. At $d=$ 750 nm thickness (the value used in our experimental prototype) the absolute transmission is low around 0.1, but $\Delta T/T$ is maximum. The absolute modulation depth, instead, (Figure \ref{fig:A6}b), is maximum at $d=$ 450 nm and slightly red-shifted if compared to the experimental device. A device with $d=$ 550 nm would reach $\Delta T/T = 10 \%$  at an absolute transmission of 0.15, which would place the device as one of the first usable transmission-mode modulators for mid-infrared free-space telecommunications. 


\section{Nonlinear modulation}

\begin{figure*}
    \centering
    \includegraphics[width=1\textwidth]{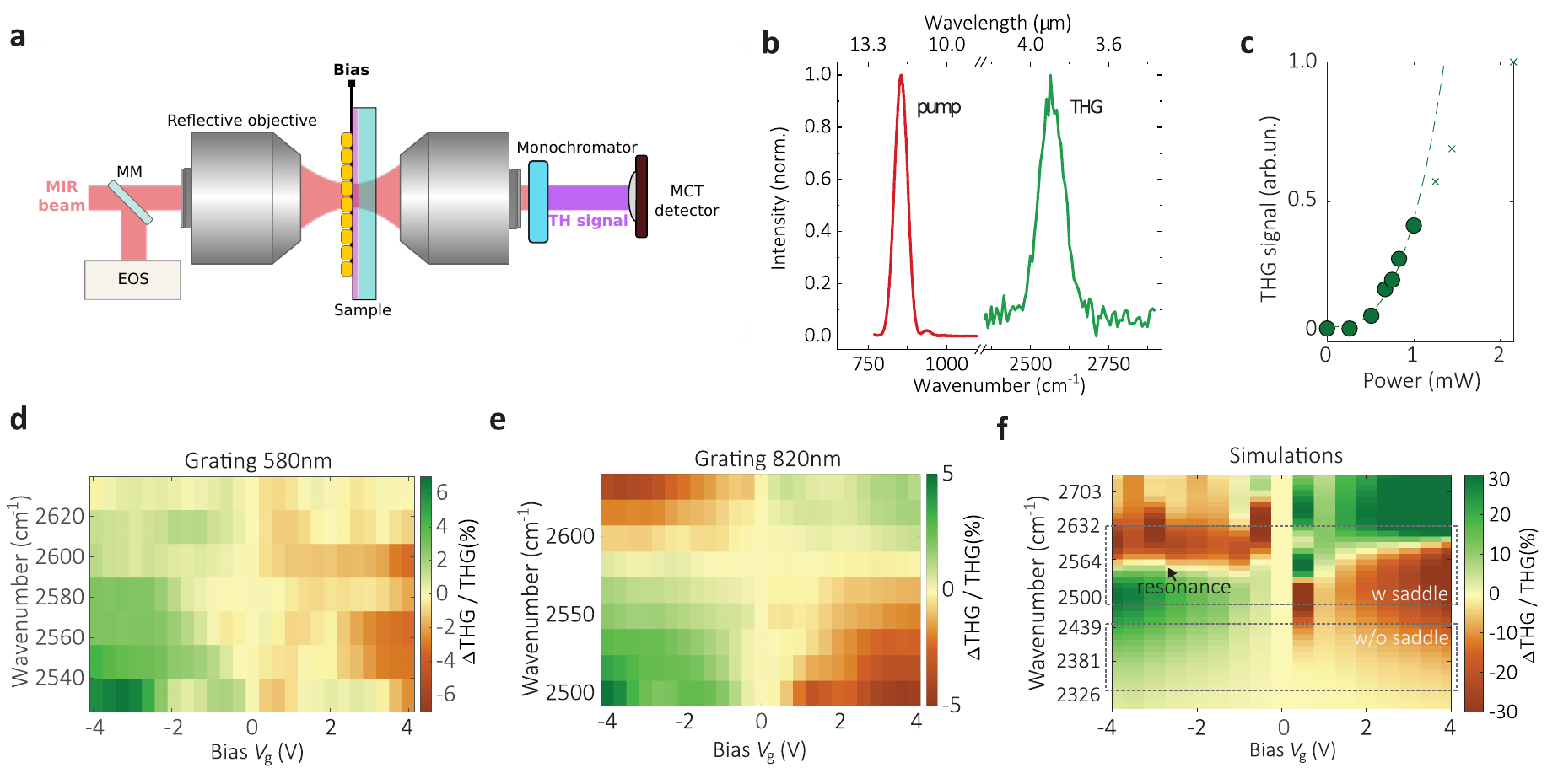}
    \caption{\textbf{Gate-modulation of the nonlinear response.} (a) Sketch of the THG setup. (b) MIR pump spectrum measured with electro-optic sampling (FF and relative TH spectrum measured with a monochromator. (c) Power $P$ dependence of the THG signal showing $P^3$ power law, followed by saturation at high $P$. (d,e) Color plots of the gate-dependent THG emission change for $w =$ 580 nm and $w=$ 820 nm respectively. (f) Simulation of the THG emission change with $V_g$. 
}   
    \label{fig:4}
\end{figure*}

We now investigate the modulation of the nonlinear optical response. We measured the third-harmonic generation (THG) from the device while varying the gate bias under conditions identical to those used in the linear experiments. As demonstrated in our previous work, the InGaAs layer exhibits a nonlinear optical response arising from free-electron density gradients at the interface \cite{rossetti2025control}. Based on this mechanism, we expect that charge density gradients induced by the field effect will lead to a significant modulation of the optical nonlinearity.

To drive the THG experiment we employ a tunable femtosecond mid-IR pump, based on difference frequency generation (DFG) between the fundamental wavelength of a Yb:KGW laser (1030nm) and the output of a nonlinear optical parametric amplifier (NOPA) with a center wavelength of 1129 nm. The mid-infrared pulses are focused onto the sample with a reflective objective, and the THG emission is measured in transmission mode. The mid-IR pump is spectrally suppressed using a sapphire shortpass filter and a monochromator. The detector is a nitrogen-cooled MCT. A schematic of the experimental setup is shown in Fig. \ref{fig:4}(a). 

In Figure \ref{fig:4}(b) we plot the spectrum of the mid-IR pump beam (red curve), acquired via electro-optic sampling, and the relative THG spectrum. The pump pulses are tuned to $\omega_{\mathrm{pump}}= 830$ cm$^{-1}$, which is resonant with the first-order $\omega_{\mathrm{res}}$ at $w$ = 820 nm. Figure \ref{fig:4}c shows a representative power dependence of the THG, reported here for $w=$ 820 nm. The cubic power-law behavior observed at low pump power confirms that the emission originates from a THG process. At higher pump powers, a saturation behavior is observed, which may be attributed to free-electron heating \cite{tielrooij2022milliwatt}. As discussed in our previous work \cite{rossetti2025control}, the optical nonlinearity responsible for THG may arise from the nonlinear hydrodynamic response of the free electrons. However, additional contributions to the nonlinear response may also be present, including the bulk dielectric $\chi^{(3)}$ of both the HfO$_2$ and InGaAs layers, as well as free-electron heating in a non-parabolic bandstructure. Although the intrinsic values of $\chi^{(3)}$ are relatively small, the effective nonlinear conversion efficiency can be significantly enhanced by the strong electric field confinement in the 1D-patch cavity.

Regardless of the microscopic origin of the optical nonlinearity, we demonstrate its modulation by an applied voltage. Figures \ref{fig:4}d-e show the voltage-induced variation of the THG emission in 2D-colorplots. The differential THG is defined as $\Delta THG/THG = \frac{THG - THG(V=0)}{THG(V=0)}$. For both gratings, we observe a significant modulation of the emission controlled by the applied voltage. The modulation depth is around 10\% in both gratings.

For $w$ = 580 nm (Figure \ref{fig:4}d), we observe that the THG emission increases under negative bias (toward depletion) and decreases under positive bias (towards electron accumulation).  This behavior is consistent with the frequency shift of the plasmonic mode shown in Fig. \ref{fig:3}: $\omega_{\mathrm{pump}}$ lies on the red side of the plasmonic resonance for $V_g=0$ and, under negative bias, $\omega_{\mathrm{res}}$ redshifts. Conversely, the THG efficiency decreases under positive voltages because $\omega_{\mathrm{res}}$ blueshifts away from $\omega_{\mathrm{pump}}$. Moreover, the maximum modulation occurs at the maximum slope of the plasmonic lineshape, whereas a node appears when $\omega_{\mathrm{res}} \sim \omega_{\mathrm{pump}}$.
In Figure \ref{fig:4}e the THG modulation for $w$ = 820 nm shows a saddle point instead of a node where $\omega_{\mathrm{res}} \sim \omega_{\mathrm{pump}}$. When $\omega_{\mathrm{res}} > \omega_{\mathrm{pump}}$ the behavior is similar to $w= 580$ nm, but when the $\omega_{\mathrm{res}} < \omega_{\mathrm{pump}}$ the $V_g$-dependence is reversed. 

To support the qualitative interpretation discussed above, we perform numerical simulations (all technical details are reported) in the SM). Figure \ref{fig:4}f shows the calculated response for $w = 685$ nm. The simulations reproduce the main experimental features, however the calculated modulation depth is significantly larger than the measured one. This discrepancy may arise from inhomogeneous broadening of the plasmonic resonance in the fabricated device: even small deviations can be amplified in the nonlinear model, leading to substantial quantitative differences. This simulation result indicates that improvement in the fabrication process may lead to substantially higher modulation depths.

\section{Discussion}

 The presented device allows for the formation of localized plasmons in the gap below the metal fingers and the InGaAs layer, leading to a good spatial overlap with the semiconductor region sensitive to the applied $V_g$ that in turn tunes $\omega_{\mathrm{res}}$, effectively modulating the linear absorption and the nonlinear emission of the device. The electric field vector in the field-effect gate configuration reaches its maximum value at the interface between the gate oxide (HfO$_2$) and the doped semiconductor (n$^+$ InGaAs). The demonstrated active electrical control of nonlinear properties of this interface opens the door to a new range of photonic applications such as optical mixers for free-space optical communications in the mid-IR, as already shown in previous works on quantum well modulators \cite{BelkinNonlinear}, and optical neural networks with programmable nonlinear activation functions \cite{rossetti2025control, DelucaPRL}. 
 
 The modulation depth can be further increased by increasing the charge density modulation at the semiconductor interface. This could be achieved either by reducing the defect state density in HfO$_2$ through surface treatments, or by employing a different gate oxide material such as Al$_2$O$_3$, which is widely used in InGaAs-based field-effect devices \cite{krylov2015physical,eom2020development}.   Another promising route to enhance the modulation depth consists in exploiting the electrically tunability of the optical phase of plasmonic resonances \cite{BelkinLinear, shaltout2019spatiotemporal}. Near $\omega_{\mathrm{res}}$, small carrier-density variations can induce significant optical phase shifts. Such phase modulation could be converted into intensity modulation using a Mach–Zehnder configuration \cite{smolyaninov2019, alvarez2025ultrahigh}. This approach would enable significantly larger on/off ratios while preserving the intrinsic speed and broadband operation of the device, both in the linear and nonlinear regimes.

Crucially, by reducing the metal patch footprint, hence total capacitance of the grating-gate, the modulation speed of the device may enter the GHz range. For a total device area of 50 $\times$ 50 $\mu$m$^2$ and a filling factor of 0.25 ($w \approx$ 500 nm, $p = 2 \mu$m), the calculated device capacitance $C$ gives a cutoff frequency $f_{-3\mathrm{dB}} = \frac{1}{2\pi R C} \approx 1$  GHz, with $R = 50$ $\Omega$ taken as the output resistance of a typical radio-frequency synthesizer.

\section{Conclusion}
We have presented a mid-IR plasmonic electro-absorption modulator based on free carrier density tuning able to modulate its linear and nonlinear responses around $\lambda = 12\mu$m. The application of a gate voltage changes the frequency of the plasmonic absorption band of 30\% of its linewidth. Moreover, the third-harmonic emission intensity has a gate-modulation depth of 10\%. These values could be improved with further device and material design to meet competitive performances with other mid-IR modulators. Our free-space modulator is based on simple heavily-doped semiconductor films, in which epitaxial layer thickness constraints are not stringent. The present electron-doped InGaAs material could easily operate across the entire 8–13 $\mu$m spectral range, also corresponding to the free-space telecom-relevant mid-IR atmospheric transparency window, simply by modifying the design of the plasmonic 1D-patch cavity.
In perspective, the modulation of the linear and nonlinear optical response in heavily-doped-semiconductor plasmonic devices will allow to design reconfigurable optical elements for free-space and integrated mid-IR photonics.

\begin{acknowledgments}
This work was supported by the European Union under the European Innovation Council (EIC) project NEHO (Grant No.101046329). Views and opinions expressed are those of the authors only and do not necessarily reflect those of the European Union or the EIC and SMEs Executive Agency (EISMEA). Neither the European Union nor the granting authority can be held responsible for them. 
We acknowledge cleanroom help from E. Herth and GL Ngo. This work was also partly supported by the RENATECH network and the General Council of Essonne. R.C. acknowledges partial support from the ERC Advanced Grant “SMART-QDEV”
\end{acknowledgments}


\bibliography{references}

\end{document}